\documentclass[conference]{IEEEtran}
\IEEEoverridecommandlockouts
\usepackage{cite}
\usepackage{amsmath,amssymb,amsfonts}
\usepackage{algorithmic}
\usepackage{graphicx}
\usepackage{url}
\usepackage{textcomp}
\usepackage{stfloats}
\usepackage{url}
\usepackage{verbatim}
\usepackage{graphicx}
\usepackage{caption}
\usepackage{subcaption}
\usepackage[ruled, vlined, linesnumbered]{algorithm2e}
\usepackage{cite}
\usepackage{booktabs}
\usepackage{amssymb,amsfonts,bm}
\usepackage{amsmath,tikz}
\usepackage{color}
\usepackage{mathtools}
\usepackage[hidelinks]{hyperref} 
\usepackage{rotating}
\usepackage{blkarray}
\usepackage{physics}
\usetikzlibrary{arrows}
\usepackage{makecell}

\usepackage{amsthm}
\usepackage{comment}
\usepackage{multirow}
\usepackage{geometry}
\usepackage{textcomp}
\usepackage{xcolor}
\def\BibTeX{{\rm B\kern-.05em{\sc i\kern-.025em b}\kern-.08em
    T\kern-.1667em\lower.7ex\hbox{E}\kern-.125emX}}
\begin{document}

\title{TAME: Temporal Audio-based Mamba for Enhanced Drone
Trajectory Estimation and Classification\\
{}
\thanks{This study was supported in part by the Postgraduate Research \& Practice Innovation Program of Jiangsu Province(KYCX24\_0612)}
}


\author{
\IEEEauthorblockN{
Zhenyuan Xiao
}
\IEEEauthorblockA{
\textit{College of Automation Engineering,} \\
\textit{Nanjing University of Aeronautics and Astronautics}\\
Nanjing, China \\
zy.xiao@nuaa.edu.cn
}
\vspace{10pt}
\IEEEauthorblockN{
Guili Xu}

\IEEEauthorblockA{
\textit{College of Automation Engineering,} \\
\textit{Nanjing University of Aeronautics and Astronautics}\\
Nanjing, China \\
Corresponding Author: guilixu2002@163.com
}
\and
\IEEEauthorblockN{
Huanran Hu}

\IEEEauthorblockA{
\textit{College of Automation Engineering,} \\
\textit{Nanjing University of Aeronautics and Astronautics}\\
Nanjing, China \\
sz2203033@nuaa.edu.cn
}
\vspace{10pt}
\IEEEauthorblockN{
Junwei He}
\IEEEauthorblockA{
\textit{College of Automation Engineering,} \\
\textit{Nanjing University of Aeronautics and Astronautics}\\
Nanjing, China \\
junwei\_he0618@nuaa.edu.cn
}
}

\maketitle

\begin{abstract}
The increasing prevalence of compact UAVs has introduced significant risks to public safety, while traditional drone detection systems are often bulky and costly. To address these challenges, we present TAME, the Temporal Audio-based Mamba for Enhanced Drone Trajectory Estimation and  Classification. This innovative anti-UAV detection model leverages a parallel selective state-space model to simultaneously capture and learn both the temporal and spectral features of audio, effectively analyzing propagation of sound. To further enhance temporal features, we introduce a Temporal Feature Enhancement Module, which integrates spectral features into temporal data using residual cross-attention. This enhanced temporal information is then employed for precise 3D trajectory estimation and classification. Our model sets a new standard of performance on the MMUAD benchmarks, demonstrating superior accuracy and effectiveness. The code and trained models are publicly available on GitHub \url{https://github.com/AmazingDay1/TAME}.
\end{abstract}

\vspace{-0.5em}
\section{Introduction}
\label{sec:intro}

\begin{figure}[t]
\centering
\includegraphics[width=7.4cm]{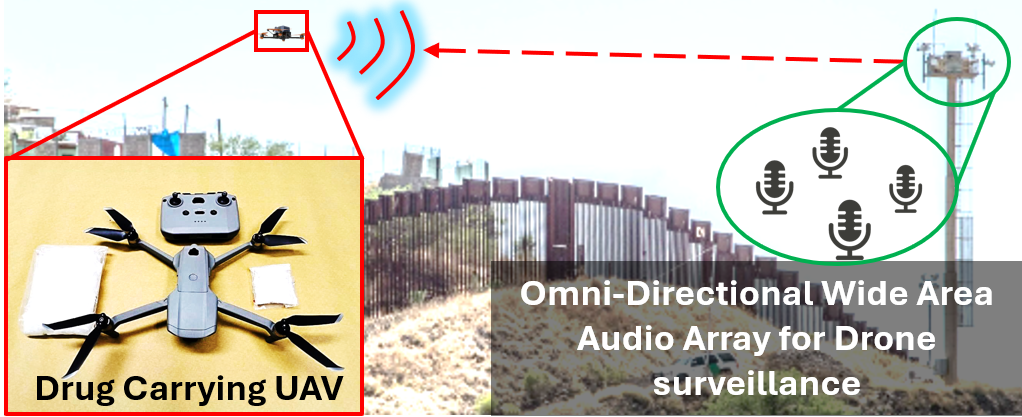}
\vspace{-0.5em}
\caption{Motivation of our proposed solution.}
\label{fig0} 
\vspace{-2.5em}
\end{figure}

\indent \hspace{13pt} Compact unmanned aerial vehicles (UAVs) \cite{esfahani2019towards} are popular for their portability, ease of use, and affordability, making them valuable in fields such as transportation \cite{cao2023neptune,cao2023path}, photography \cite{liu2024distance,xu2024cost,liu2023non,lyu2023spins},  search and rescue \cite{yuan2021survey,cao2021distributed,yuan2024largescale,10802691}. However, these same features also make them difficult to detect \cite{10611652} when used maliciously, posing threats to air traffic control and security and being exploited in warfare and border drug smuggling \cite{cao2022direct}, as shown in Fig. \ref{fig0}.

Existing anti-UAV detection methods often rely on single-modality techniques~\cite{zhao21082nd_IF,zhao20233rd_IF,munir2024investigation_rgb_dtc,zheng2021air_rgb_dtc,liang2024unsuperviseduav3dtrajectories,lei2024audioarraybased3duav,liang2024separatingdronepointclouds}, such as radio signals \cite{yuan2024large,nguyen2024uloc}, radar, or images \cite{yuan2014autonomous}, but these approaches can be error-prone, particularly due to limitations in radar cross section~\cite{deng2024multi_point,wu2024vehicle_point,vrba2023onboard_point} or pixel size~\cite{lyu2023real_IF_dtc}. While some methods attempt to overcome these issues through multi-modal fusion~\cite{vora2023dronechase, yang2024av_Audio-visual,xiao2024avdtec,cao2024mopa}, combining inputs like multi-spectral images and audio data, they often rely on unrealistic assumptions~\cite{huang2023anti_IF_tracking}, such as perfect targeting or high vantage points, making them less practical in real-world scenarios. Audio, however, presents a distinct advantage by providing reliable information about the direction~\cite{wang2022large_audio_cls,ali2024exploitinge_audio_cls}, classification~\cite{erol2024audio,shams2024ssamba}, and distance of sound sources~\cite{yang2023av_audio_img_fusion, tao2021someone}, which is largely unaffected by environmental conditions. This is particularly useful given the significant noise generated by UAVs, making them easier to detect through audio~\cite{vora2023dronechase}. Recognizing these benefits, recent studies~\cite{yang2024av_Audio-visual} have begun to advocate for integrating audio-based solutions to improve UAV detection accuracy.

In response to these challenges, we propose a novel anti-UAV model using audio, called Temporal Audio-based Mamba \cite{erol2024audio} for Enhanced Drone Trajectory Estimation and Classification (TAME), which utilizes the selective state-space model (SSM) ~\cite{gu2023mamba} to effectively process temporal sequences in audio data for UAV detection. This model enables omnidirectional 3D spatial area detection, addressing the limitations of current methods that are primarily focused on classification, tracking, and position estimation using visual data. While effective, traditional multi-modal fusion approaches demand high computational power, which limits their feasibility for mobile or wearable applications. By leveraging the efficiency of SSM, TAME offers a more practical solution for accurate and efficient UAV detection, especially in challenging real-world environments.
Our contributions are summarized as follows:

\begin{enumerate}
  \item We propose the first temporal-spectral mamba for feature extraction, using learnable patches to capture and enhance differential features, effectively mapping the temporal and spectral dynamics of sound propagation.
  \item We propose the Temporal Feature Enhancement Module, which integrates spectral and temporal audio features with cross-attention and uses residual connections and learnable patches to enhance trajectory estimation and classification.
  \item We benchmarked against SOTA methods, achieved the best results in all challenging conditions, and made our solution open-source.
\end{enumerate}

\begin{figure*}[t]
\centering
\centering
\includegraphics[width=0.95\linewidth]{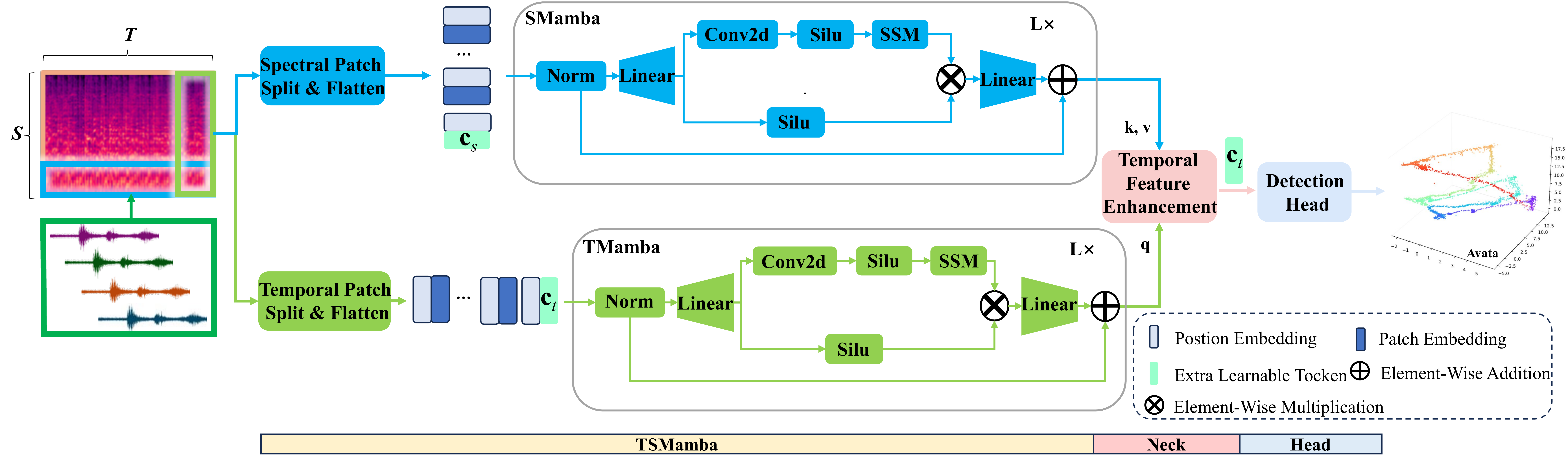}
\vspace{-1.1em}
\caption{Proposed TAME Architecture for audio-only UAV detection.}
\label{fig1} 
\vspace{-2em}
\end{figure*}

\vspace{-5pt}
\section{Method}

\indent \hspace{13pt} In this section, we introduce TAME, an end-to-end architecture for UAV trajectory estimation and classification using audio. The overall architecture is illustrated in Fig. \ref{fig1}. TAME consists of the temporal spectral mamba (TSMamba) backbone, the temporal feature enhancement (TFE) neck, and the detection head. TSMamba takes the mel-spectrogram of audio as input, extracts temporal and spectral features using Temporal Mamba (TMamba) and Spectral Mamba (SMamba), respectively, and integrates these features through TFE, before sending them to the detection head for final output. 
TMamba extracts the temporal difference of arrival as a feature, while SMamba captures spectral attenuation as a feature.
\vspace{-.8em}
\subsection{Temporal Spectral Mamba }    \label{sec3.A}
\indent \hspace{13pt} Audio features are extracted by TSMamba, as shown in Fig. \ref{fig1}. 
Let ${*}$ denote the temporal (\textit{t}) or spectral (\textit{s}) axis. Denote $\kappa$ as the number of microphone channels, \textit{R} as the
temporal width of the spectrogram, \textit{S} as the spectral height, \textit{W} is the width of the patch, and \textit{H} as the height of the patch.
To extract audio features, multichannel audio is converted into mel-spectrograms $\textbf{$\chi$}\in \mathbb{R}^{\kappa\times \textit{R}\times \textit{S}}$, which serve as input to our model. As the standard Mamba is a 1D sequence model, the spectrogram is split and flattened along the temporal and spectral axis to obtain $\textbf{p}_{*}$, which consists of the temporal patch sequence $\textbf{p}_{t }\in \mathbb{R}^{{J}\times \textit{($\kappa$WS)}}$ and the spectral patch sequence $\textbf{p}_{s }\in \mathbb{R}^{{J}\times \textit{($\kappa$RH)}}$. Inspired by ViT~\cite{dosovitskiy2020image_transformer}, we append learnable tokens $\textbf{c}_{* }$ in both temporal and spectral directions. These tokens are used by selective SSM to summarize the temporal difference and spectral attenuation features of the entire patch sequence. To use those tokens,  $\textbf{p}_{* }$ is linearly projected to a vector of size \textit{D} with added position embeddings $\textbf{E}_{pos}\in \mathbb{R}^{{(J+1)}\times \textit{D}}$. With the linearly projected vector, we can generalize the overall feature vector $\textbf{X}_{*}$:
\begin{equation} 
\textbf{X}_{*}=[\textbf{p}_{\ast}^{1}\textbf{W};\textbf{p}_{\ast}^{2}\textbf{W};...;\textbf{p}_{\ast}^{J}\textbf{W};\textbf{c}_{*}]+\textbf{E}_{pos}
\end{equation}
\hspace{13pt} Non-overlapping convolutions are employed for patch splitting. For the temporal patch split (horizontal), the height is kept the same as the spectrogram, and TMamba scans from left to right along the temporal axis to extract the temporal difference of the arrival, capturing sound propagation feature. For the spectral patch split (vertical), the width is kept the same as the spectrogram, and SMamba scans from top to bottom along the spectral axis to extract global spectral attenuation features. Both TMamba and SMamba are composed of selective SSM. The selective SSM fuses audio features from the previous patch with those of the current patch. 
\vspace{-.8em}
\begin{figure}[t]
\centering
\includegraphics[width=8.6cm]{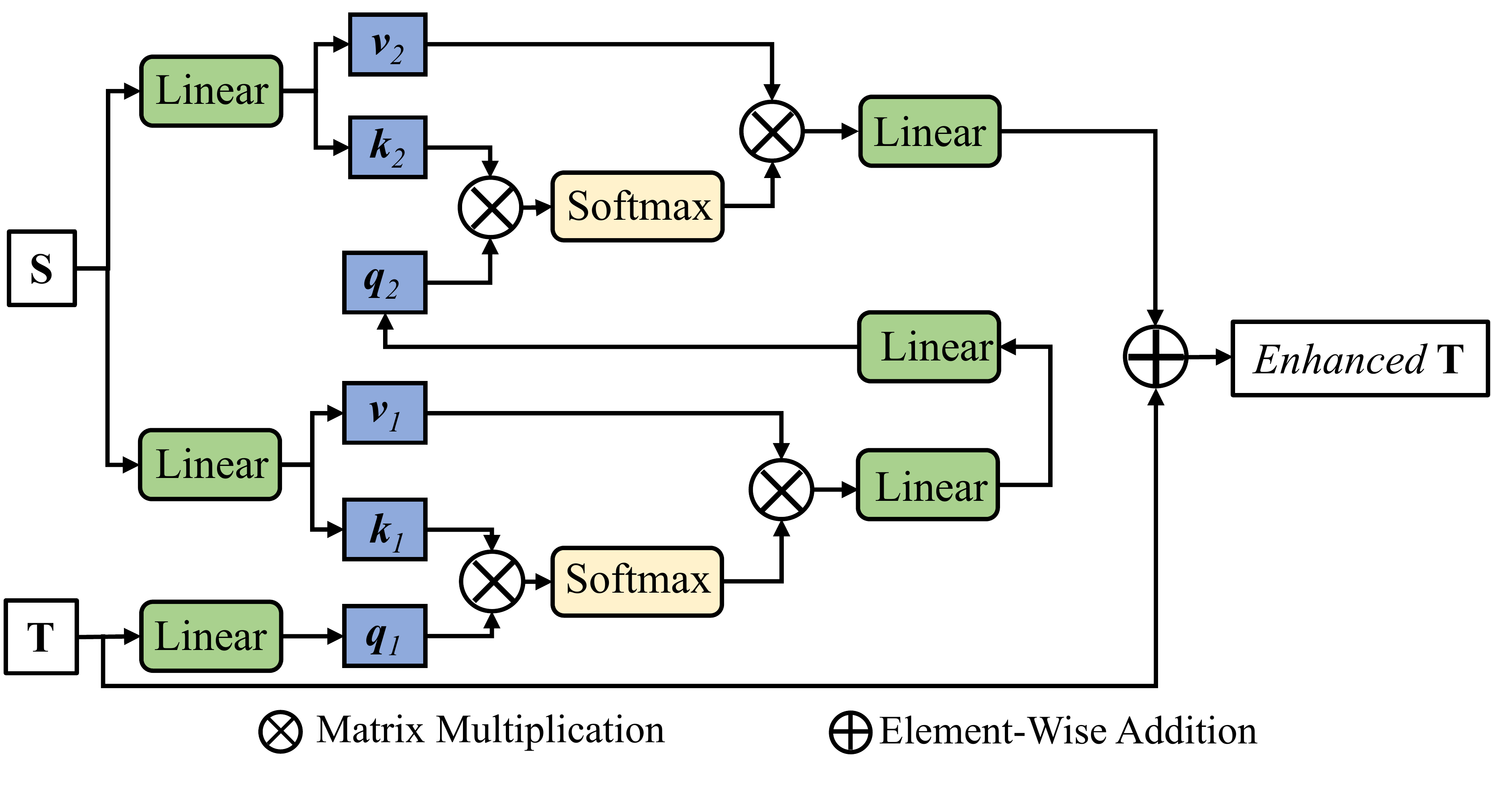}
\vspace{-2.8em}
\caption{Temporal Feature Enhancement Module}
\label{fig3} 
\vspace{-2em}
\end{figure}

\subsection{Temporal Feature Enhancement}   \label{sec3.B}
\indent \hspace{13pt} In the TSMamba feature backbone, differences between patch features are filtered.  The propagation time of sound correlates with distance. When scanning each patch from left to right, different spectrograms exhibit varying degrees of attenuation at the same distance. Thus, temporal features encode both location and some category information. While the spectral features capture global spectral ratios over time and also encode category information. 
Simultaneously using both features can lead to information redundancy, negatively impacting the final outcome.
Therefore, we design a Temporal Feature Enhancement (TFE) module, as shown in Fig.  \ref{fig3}. This module employs cross-attention to integrate spectral features into temporal features, enabling UAV classification and trajectory estimation solely based on temporal features. The TFE calculation is as follows:
\vspace{-.4em}
\begin{equation} 
\begin{matrix}
\begin{aligned}

\text{TFE}(\textbf{T},\textbf{S})=\textbf{T} +  \text{Attention}({{q}}_{2},\textbf{S}{\textbf{W}}^{{k}_{2}},\textbf{S}{\textbf{W}}^{{v}_{2}}){\textbf{W}}^{}

 \\ {q}_{2}= \text{Attention}(\textbf{T}{\textbf{W}}^{{q}_{1}},\textbf{S}{\textbf{W}}^{{k}_{1}},\textbf{S}{\textbf{W}}^{{v}_{1}})

 \\\text{Attention}(\textbf{Q},\textbf{K},\textbf{V})=\text{softmax}(\frac{\textbf{Q}{\textbf{K}}^{\text{T}}}{\sqrt{{d}_{k}}})\textbf{V}
\end{aligned}
\end{matrix}
\end{equation}

where $\textbf{W}\in {\mathbb{R}}^{{d}_{m}\times {d}_{k}}$  represents learnable parameter matrices. $\textbf{T}\in {\mathbb{R}}^{{(J+1)}\times {d}_{m}}$ represents temporal features, and $\textbf{S}\in {\mathbb{R}}^{{(J+1)}\times {d}_{m}}$ represents spectral features. Meanwhile, \textbf{Q}, \textbf{K} and \textbf{V} are divided into ${n}$ attention heads.

\subsection{Detection Head}  \label{sec3.C}
\indent \hspace{13pt} First, the learnable temporal token $\textbf{c}_{t }$ is extracted from the enhanced temporal features. The token is then sent to a detector for UAV trajectory estimation and classification. The detection head consists of two heads: a trajectory prediction head and a drone classification head. Both heads utilize multi-layer perceptrons (MLPs) to map outputs.

\textbf{Trajectory prediction head:} UAV audio often contains various noises that can result in inaccurate trajectory predictions. To ensure model stability against noise during training, we use L1 loss. This approach prevents the model from deviating from the expected training trajectory due to noise. The loss function is defined as follows:

\begin{equation} 
\begin{matrix}
\begin{aligned}
{L}_{pos}=\frac{1}{N}\displaystyle\sum_{i=1}^{N}\left | \hat{{O}_{i}}-{o}_{i}\right |,
\end{aligned}
\end{matrix}
\end{equation}

where \textit{N} denotes the total number of UAV trajectory, $\hat{O}$ represents the ground truth 3D trajectory, and ${o}$ denotes the predicted trajectory.

\begin{table*}[t]
\centering
\footnotesize
\caption{3D trajectory estimations and accuracy for MMAUD V1 Dataset \cite{yuan2024mmaud}.}
\vspace{-6pt}
\label{tab1}
\renewcommand{\arraystretch}{1.5}

\begin{tabular}{cccccccccccccccc}

\toprule 
\hline

\multirow{2}{*}{Modality}& \multirow{2}{*}{Network}& \multirow{2}{*}{Params}& \multicolumn{5}{c}{Light}  & \multicolumn{5}{c}{Dark} & \multirow{2}{*}{$\overline{\text{APE}}$} &
 \multirow{2}{*}{$\overline{\text{Acc}}$(\%)}\\
 
\cmidrule(lr){4-8} \cmidrule(lr){9-13}
& &(M) &  $D_x$ & $D_y$ & $D_z$ &$\text{APE}$& $\text{Acc(\%)}$ & $D_x$ & $D_y$ & $D_z$ &$\text{APE}$ & $\text{Acc(\%)}$\\
\midrule

\multirow{2}{*}{Visual} & VisualNet~\cite{yang2023av_audio_img_fusion} & 31.18& 0.24 &  {0.39} & \underline{0.32} & {0.65} & \underline{99.7 } & 1.98  & 6.10 & 8.13  & 11.45 & 11.3 & 6.05 & 55.5 \\ 

 & DarkNet~\cite{bochkovskiy2020yolov4} & 40.69 & 0.23  & 0.46  & \textbf{0.23}  &  \underline{0.63}  & \textbf{100}  & 1.84   & 5.50  & 4.57 & 8.31  & 25.9 & 4.47 & 63.0  \\

\hline

 Audio & TalkNet~\cite{tao2021someone}& 15.80 & 0.31  & 0.69  & 0.44  & 0.99  &  \textbf{100}  & 1.13  & 3.39 & 3.92  & 5.82 & 47.4 & 3.41 &73.7  \\

Visual & AV-PED~\cite{yang2023av_audio_img_fusion}& 32.36 &  0.31 & 0.50  & 0.59  & 0.97  & 98.5  & 0.58   & 1.54  & 2.26 & 3.13  & 80.7 & 2.01&89.6  \\

Fusion & AV-FDTI ~\cite{yang2024av_Audio-visual}& 92.28 &\underline{ 0.13}  &  \underline{0.31} & 0.38  & 0.58  & 99.6   & \underline{0.35}  & \underline{1.06} & \underline{1.10}  & \underline{1.89}& \underline{88.3 }&\underline{ 1.24} &\underline{94.0}  \\
\hline

\multirow{3}{*}{Audio}& AudioNet ~\cite{yang2023av_audio_img_fusion}& 1.45 & 0.60   & 1.76   & 1.59  & 2.80  &  79.8  & 0.60  &1.76  &1.59   & 2.80&79.8 & 2.80& 79.8  \\

 & DroneChase~\cite{vora2023dronechase}& 0.24 & 0.54  & 1.59  & 1.51  & 2.64  & 80.6  & 0.54  & 1.59 & 1.51  & 2.64 & 80.6 & 2.64 & 80.6  \\

&\textbf{TAME}(Ours) & 6.61 &\textbf{0.11 }  & \textbf{0.30}  & 0.34  &\textbf{0.55 }  & 98.0  &\textbf{ 0.11}  & \textbf{0.30}  & \textbf{0.34}  &\textbf{0.55}   & \textbf{98.0}  &\textbf{0.55}   & \textbf{98.0}   \\
    \hline
    \bottomrule
\end{tabular}
\\
\footnotesize{Best results are highlighted in $\textbf{bold}$, and second best in $\underline{underline}$. This notation apply for rest of the paper. $\overline{Overline}$: the mean of day and night.}
\vspace{-2em}
\end{table*}

\textbf{Classification prediction head:} Attributes of the UAV, such as size and audio, are indicative of its category. It is crucial for anti-UAV systems. Classification information enhances the system’s ability to pinpoint the UAV’s 3D coordinates. Additionally, the anti-UAV system can take appropriate actions based on the UAV type. Therefore, a classification head is designed to perform UAV classification, using cross-entropy loss, defined as follows:

\begin{equation} 
\begin{matrix}
\begin{aligned}
{L}_{cls}=-\frac{1}{N}\displaystyle\sum_{i=1}^{N}{y}_{i}log({p}_{i}),
\end{aligned}
\end{matrix}
\end{equation}

where $N$ denotes the total number of UAV type, ${y}_{i}$ represents the ground truth class, and ${p}_{i}$ represents the predicted class.

Therefore, the overall training loss function is given by

\begin{equation} 
\begin{matrix}
\begin{aligned}
{L}_{total} =  {L}_{cls} + \gamma {L}_{pos},
\end{aligned}
\end{matrix}
\end{equation}

where $\gamma$ is the balancing factor for the multi-task loss.

\begin{figure*}
\centering
\includegraphics[width=0.99\linewidth]{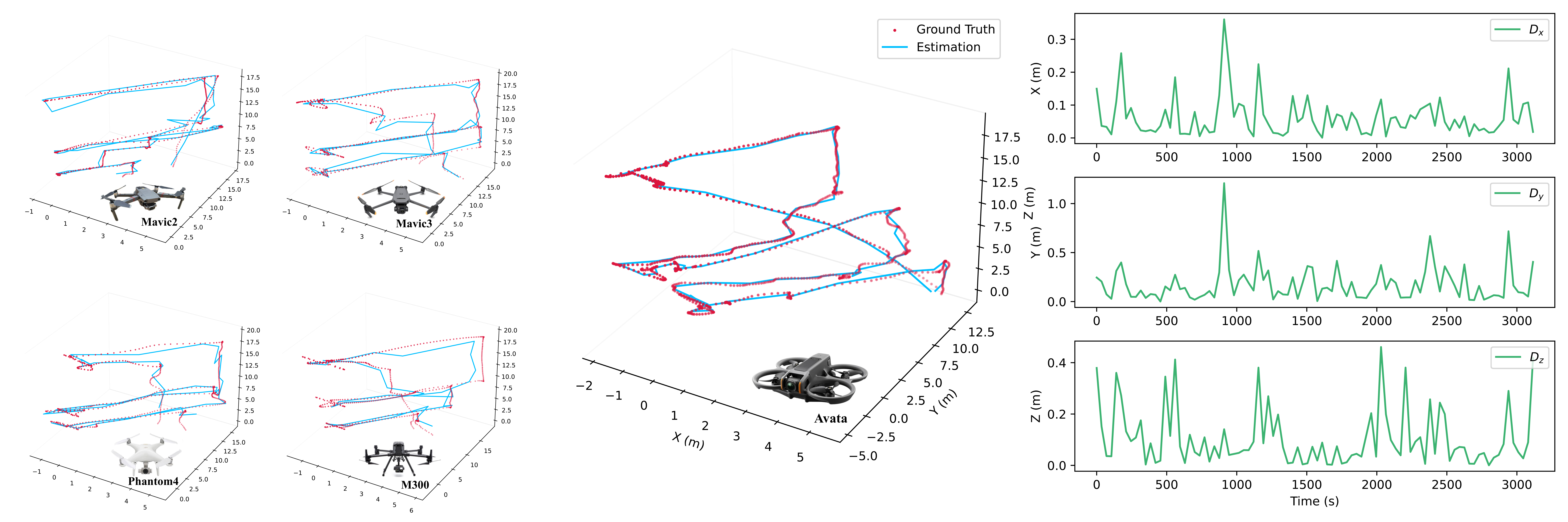}
\caption{Test set trajectory estimation: Red curves represent ground truth, blue curves show predicted trajectories.}
\label{fig4} 
\end{figure*}

\begin{figure}
\centering
\includegraphics[width=7.8cm]{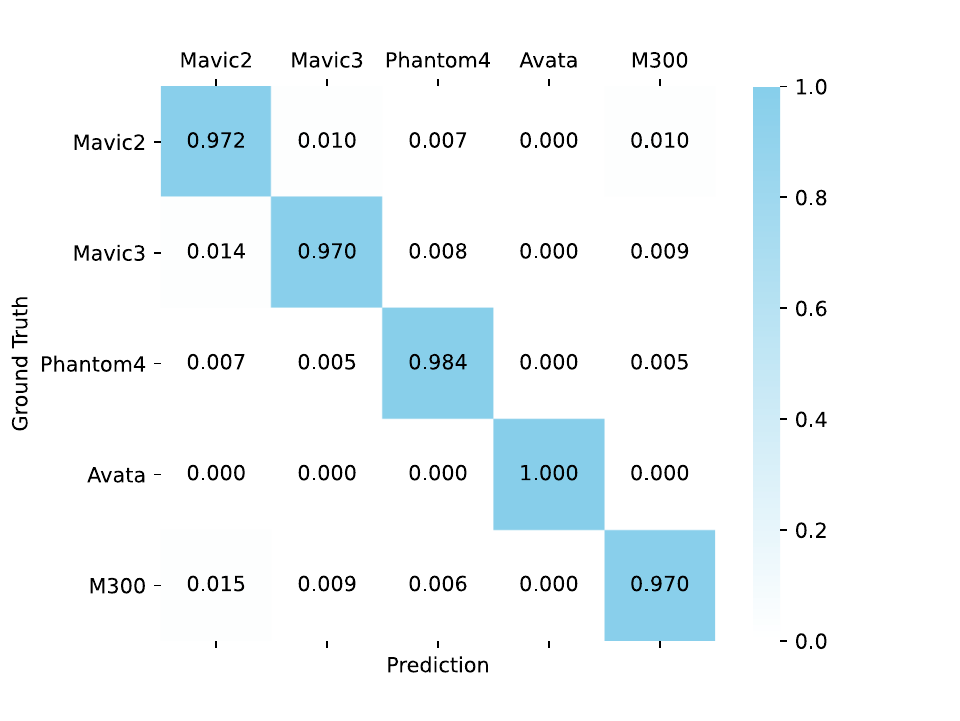}
\vspace{-1.5em}
\caption{The confusion matrix for the classification results.}
\label{fig5} 
 \vspace{-2em}
\end{figure}

\section{EXPERIMENT}

\subsection{Dataset} \indent \hspace{13pt} We use the MMAUD anti-UAV dataset~\cite{yuan2024mmaud}, containing images, lidar, millimeter-wave point clouds, four-channel audio, and ground truth. The UAV flight range is 7 m × 25 m × 22 m. Audio data, sampled at 48 kHz, is segmented into 0.2-second intervals and converted to mel-spectrograms (224 × 16 resolution). The dataset is split 7:3 for training and testing.

\subsection{Experimental Setting} \indent \hspace{13pt} 
 \textbf{Implementation Details:} The model is trained on an NVIDIA RTX 3090 GPU using Adam, with a batch size of 64, a learning rate of 0.0001, and 200 epochs. For patch splitting, \textit{J}=16, \textit{W}=4, \textit{H}=1; for TSMamba, \textit{L}=12;  for TFE, ${n}$ = 6, ${d}_{m}$ = ${d}_{k}$ = 192. The multi-task balance factor is $\gamma$=2. Brightness attenuation is applied during training/testing.

\textbf{Evaluation Metrics:} We use the L1 norm for center distances (${D}{x}$, ${D}{y}$, ${D}_{z}$) and average trajectory error (APE) for trajectory estimation, and accuracy (Acc) for classification performance.

\vspace{-10pt}
\subsection{Baseline Selections} \indent \hspace{13pt} We compare our model against several anti-UAV approaches using visual\cite{bochkovskiy2020yolov4,yang2023av_audio_img_fusion}, audio\cite{vora2023dronechase,yang2023av_audio_img_fusion}, and audio-visual fusion\cite{yang2023av_audio_img_fusion,yang2024av_Audio-visual,tao2021someone}. The baselines include self-supervised, active speaker detection, pedestrian detection, and multi-task models, focusing on different modalities and feature extraction methods.

\subsubsection{Trajectory Estimation and UAV Classification}
\indent \hspace{13pt} Table \ref{tab1} shows that TAME achieves state-of-the-art results in both $\overline{\text{APE}}$ and $\overline{\text{Acc}}$, outperforming audio, visual, and audio-visual fusion methods, especially at night. TAME's ability to capture global temporal features in audio spectrograms gives it an edge over other models like DroneChase and AudioNet. While visual-based models excel during the day, their performance declines at night, and audio-visual fusion methods struggle with lighting changes. Although attention mechanisms in AV-PED and AV-FDTI improve performance, they still fall short of TAME. This highlights the superiority of single-modal approaches like TAME in UAV trajectory estimation and classification. Fig. \ref{fig4} and \ref{fig5} illustrate UAV position estimates and classification results, showing that most UAVs are accurately detected, except for some cases like M300, where weak audio signals are masked by environmental noise.

\begin{table}[t]
\centering
\caption{The ablation study of TAME with different modules and feature fusion. SFE is a spectral feature enhancement.}
\label{tab2}
\renewcommand{\arraystretch}{1.5}
\begin{tabular}{lccccccc}
\hline
TMamba & SMamba  & SFE & TFE & $\text{$\overline{\text{APE}}$}$& $\text{$\overline{\text{Acc}}$(\%)}$\\
\hline
$\checkmark$ &     &  &  &{0.59} &  97.7  & \\
&$\checkmark$     &  &  & 0.87 & 95.4 &   \\
$\checkmark$& $\checkmark$    & $\checkmark$ &  & 0.68 & 96.7 &   \\
$\checkmark$& $\checkmark$    &  & $\checkmark$ & \textbf{0.55} & \textbf{98.0} &   \\

\hline
\end{tabular}
\vspace{-2em}
\end{table}
\vspace{-10pt}
\subsection{Ablation Study and Analysis}  
\indent \hspace{13pt} To verify the effectiveness of the proposed TAME, we conducted ablation experiments to assess the significance of various features and fusion methods. Table \ref{tab2} demonstrates that TAME achieves the highest overall performance in both the $\overline{\text{APE}}$ and  $\overline{\text{Acc}}$. 
This improvement is attributed to the use of TFE, which enhances temporal features.

\vspace{-5pt}
\section{Conclusion}
\vspace{-5pt}
\indent \hspace{13pt} We propose TAME, an audio-based model for detecting UAV threats that integrates spectral and temporal features for top performance in trajectory estimation and classification. Despite its effectiveness, TAME has limitations, including position estimation errors and reliance on large datasets. Future work will aim to enhance trajectory estimation and explore unsupervised methods using point cloud data.

{\footnotesize
\bibliographystyle{IEEEbib}
\bibliography{mybib}
}
\end{document}